# A computational study of expiratory particle transport and vortex dynamics during breathing with and without face masks

Ali Khosronjead[1*], Seokkoo Kang[2], Fabian Wermelinger[3], Petros Koumoutsakos[4], and Fotis Sotiropoulos[1]

[1] Civil Engineering Department, Stony Brook University, Stony Brook, NY 11794

[2] Department of Civil and Environmental Engineering, Hanyang University, Seoul 04763, South Korea

[3] Computational Science & Engineering Laboratory, ETH Zurich, CH-8092, Switzerland

[4] Institute for Applied Computational Science, Harvard University, Cambridge, MA 02138

*Corresponding author, Tel: (631) 632-9222, Email: ali.khosronejad@stonybrook.edu

**Abstract**

We present high-fidelity numerical simulations of expiratory biosol transport during normal breathing under indoor, stagnant air conditions with and without a facial mask. We investigate mask efficacy to suppress the spread of saliva particles that is underpinnings existing social distancing recommendations. The present simulations incorporate the effect of human anatomy and consider a spectrum of saliva particulate sizes that ranges from 0.1 µm to 10 µm while accounting also for their evaporation. The simulations elucidate the vorticity dynamics of human breathing and show that without a facial mask, saliva particulates could travel over 2.2 m away from the person. However, a non-medical grade face mask can drastically reduce saliva particulate propagation to 0.72 m away from the person. This study provides new quantitative evidence that facial masks can successfully suppress the spreading of saliva particulates due to normal breathing in indoor environments.

I. Introduction

The outbreak of the COVID-19 pandemic has overwhelmed health care systems around the globe, nearly devastated the global economy, and, in the United States alone, has claimed more than half a million lives. During early stages of the pandemic outbreak, health organizations such as the World Health Organization (WHO) and the United States Centers for Disease Control (CDC) recommended social distancing and mask-wearing as effective measures to prevent the spread of the disease [1-5]. The WHO and CDC social distancing guidelines specifically recommended a minimum physical separation of 1 m and 1.8 m, respectively, to limit the airborne transmission of the disease [5,6]. Researchers have sought to examine the scientific underpinning of these guidelines by investigating the underlying physics of respiratory events and transport mechanisms that contribute to the propagation of pathogen-carrying saliva droplets [6-14]. Bourouiba et al. (2014) [11], and more recently, Leung et al. (2020) [8], Fischer et al. (2020) [9], and Verma et al. (2020) [10] investigated experimentally indoor coughing and sneezing expiratory events and reported that saliva particles of such events could, under certain conditions, propagate as far as 7 m to 8 m away from the person. Tang et al. (2009) [15] used a schlieren optical method to visualize human coughing and normal breathing in a controlled experimental setting. Numerical simulations have been deployed to study various expiratory biosol transport processes. Dbouk and Drikakis [12,13] conducted Reynolds-averaged Navier-Stokes (RANS) modeling of saliva transport during coughing. Cummins et al. (2020) [14] employed a simplified numerical modeling approach to study the dispersion of multi-size saliva particles. Khosronejad et al. (2020) [16] performed high-resolution large-eddy simulation (LES) of a single cough pulse investigating the saliva plume propagation under indoor and outdoor conditions. Their study demonstrated that facial masks are effective for curbing the spread of saliva plume in indoor environments. Mittal et al. (2020) [17] introduced a mathematical framework to examine the efficacy of facial masks and estimate the risks of the airborne transition. They alluded to the importance of the range of biosol size during airborne transmission of disease and mentioned that "*it is the small (< 10 μm) particles that are likely to be entrained into the inhalation current of a person, while environmental conditions, as well as the proximity between the host and the susceptible, could allow larger particles/droplets to play a role in airborne transmission.*" Abuhegazy et al. (2020)

[18] employed an Eulerian-Lagrangian approach to model the spreading of saliva particles in a classroom setting and showed that, among others, the background airflow pattern played a crucial role in the indoor spreading of the exhaled saliva particles.

Like coughing and sneezing in which saliva particles are subject to jet-like flows exiting the mouth,[11-13,16] normal breathing produces periodic jet flows that efflux saliva particles into the ambient air downwind of the person. The flow of air-saliva mixture during coughing and sneezing has a maximum velocity of about 10 ms$^{-1}$ and is generally characterized as a turbulent flow with a Reynolds number of $Re \sim 10,000$[11], while the maximum velocity of the air-saliva mixture during normal breathing ranges from 0.5 ms$^{-1}$ to 1.0 ms$^{-1}$ which could produce a jet-flow with $Re \sim 2,000$. Although less violent than coughing and sneezing, the jet flow of normal breathing consists of three-dimensional coherent vortical flow structures responsible for much of the saliva particle transport. To the best of our knowledge, however, the vorticity dynamics of human breathing, its impact on the transport of respiratory viruses, and the effect of facial masks on curbing such spread have yet to be investigated systematically.

In this work we seek to elucidate the vortex dynamics of human breathing with and without masks and evaluate social distancing guidelines required to curb the spread of respiratory viruses. We employ an Eulerian approach to carry out high-resolution numerical simulations of multi-size saliva particles transport during normal breathing using computational grids fine enough to resolve almost all relevant scales of motion. Saliva particles in this work range in size from 0.1 μm to 10 μm, which is within the range of aerosol particulates produced during human breathing and are known to be large enough to carry the virus [1-6]. Each size class is simulated as an active scalar whose motion is governed by Boussinesq-type advection-diffusion equation with size-dependent settling velocity.[20-23] The effect of saliva evaporation, which could become significant during multiple cycles of normal breathing, is accounted for indirectly by gradually decreasing the size and settling velocity of suspended saliva particles [16,17]. As a result, saliva particles could stay in suspension for prolonged periods without settling down[17]. The three-dimensional geometry of facial coverings and the anatomy of the human body are incorporated in the model using the immersed boundary methods. The sharp interface curvilinear immersed boundary (CURVIB) method [19,20] is used to simulate the anatomy of the human face, while a novel diffused-interface immersed boundary approach is adapted to model the effect of facial masks on the flow and saliva plume transport. Normal breathing is prescribed at the opening of the mouth and the simulations are performed under stagnant air, i.e., indoor conditions. The simulations are carried out using an in-house open-source code, the Virtual Flow Simulator (VFS) code, which has been applied to a wide range of biological, environmental, and engineering turbulent flows [20-28]. Our results show that normal breathing without a facial mask generates periodic trailing jets and leading circular vortex rings that propagate forward and interact with the vortical flow structures produced in the prior breathing cycles. Such a complex vorticity field is shown to be able to transport aerosols for over 2.2 m downwind of the body. A face mask, on the other hand, is able to dissipate the kinetic energy of the breathing jet, disrupt and disorganize vortical structures and substantially limit the downwind propagation of virus-laden aerosols.

This paper is organized as follows. Section 2 describes the governing equations of fluid motion and saliva plume transport, followed by the numerical model of human anatomy and facial masks in Section 3. The computational details of the simulations are presented in Section 4. Subsequently, we discuss the simulation results in Section 5 and conclude this paper's findings in Section 6.

## II. Governing equations of fluid motion and saliva plume transport

We solve the spatially filtered continuity and Navier-Stokes equations with the Boussinesq assumption to simulate the incompressible, stratified, turbulent flow of dilute air-saliva mixture. The Navier-Stokes equations in non-orthogonal, generalized, curvilinear coordinates $\{\xi^i\}$, and compact tensor notation read as follows ($i = 1, 2$ or $3$ and $j = 1, 2,$ and $3$) [28]:

$$J \frac{\partial U^j}{\partial \xi^j} = 0 \tag{1}$$

$$\frac{1}{J}\frac{\partial U^j}{\partial t} = \frac{\xi_l^i}{J}\left[ -\frac{\partial(U^j u_l)}{\partial \xi_j} + \frac{1}{\rho_0}\frac{\partial}{\partial \xi^j}\left(\mu \frac{\xi_l^j \xi_l^k}{J}\frac{\partial u_l}{\partial \xi^k}\right) \right.$$
$$\left. -\frac{1}{\rho_0}\frac{\partial}{\partial \xi^j}\left(\frac{\xi_l^j p}{J}\right) - \frac{1}{\rho_0}\frac{\partial \tau_{lj}}{\partial \xi^j} + \frac{\bar{\rho} - \rho_0}{\rho_0} g\left(\frac{\delta_{i3}}{J}\right) + f_l \right] \tag{2}$$

where repeated indices imply summation, $\xi_l^i$ are the transformation metrics, $J$ is the Jacobian of the transformation, $U^i$ is the contravariant volume flux, $u_i$ is the Cartesian velocity component, $p$ is the pressure, $\tau_{lj}$ is the subgrid stress tensor in Large-Eddy Simulation (LES) model, $\delta_{ij}$ is the Kronecker delta, $\mu$ is the dynamic viscosity, $g$ is the gravitational acceleration, $\rho_0$ is the background density (of air), $\bar{\rho}$ is the density of the air-saliva mixture, and $f_l$ (with $l = 1, 2,$ and $3$) is the body force due to the facial mask. We employ a second-order finite differencing numerical scheme and the dynamic Smagorinsky model [29] to model the unresolved subgrid scales of turbulence in the LES. The density of the air-saliva mixture, in the term arising from the Boussinesq assumption, is calculated as [30]

$$\bar{\rho} = \rho_0(1 - \psi) + \rho_s \psi \qquad (3)$$

where $\psi$ is the volume fraction of the saliva particulates and $\rho_s$ is the density of saliva particles, which is set equal to that of the water (=1000 kg m$^{-3}$). The saliva plume concentration in the dilute mixture of air and saliva is modeled as an active scalar. We compute the concentration of saliva plume using the following convection-diffusion equation [30]:

$$\frac{1}{J}\frac{\partial(\rho_0 \psi)}{\partial t} + \frac{\partial}{\partial \xi^j}\langle \rho_0 \psi (U^j - W^j \delta_{i3}) \rangle - \frac{\partial}{\partial \xi^j}\langle (\mu \sigma_L + \mu_t \sigma_T)\frac{\xi_l^j \xi_l^k}{J}\frac{\partial \psi}{\partial \xi^k}\rangle = 0 \qquad (4)$$

where $W^j = (\frac{\xi_3^j}{J})w_s$ is the contravariant volume flux of saliva concentration in the vertical direction due to the settling velocity ($w_s$) of the particles, $\sigma_L$ is the laminar Schmidt number (=1) [31], $\sigma_T$ is the turbulent Schmidt number (=0.75) [31], and $\mu_t$ is the eddy viscosity.

### III. The numerical model of the human anatomy and facial masks

We employ the sharp-interface CURVIB method and wall model reconstruction [19,20] to model the human body's three-dimensional anatomy, as seen in Fig. 1(a). The effect of facial masks is modeled using a diffused-interface immersed boundary method. Namely, by applying a drag force on the unstructured CURVIB grid nodes used to discretize the three-dimensional geometry of the mask (Fig. 1(c)). The facial mask drag force is distributed onto the fluid nodes (Fig. 1(d)) using a smoothed discrete delta function as the kernel for transferring information between grid nodes within the fluid phase and the mask, as follows [16]:

$$f_l = \frac{1}{2}\bar{\rho}\, C_D a(u_i u_i)^{1/2} u_l \Delta(x_j - X_j) \quad (5)$$

where $C_D$ is the drag coefficient, $a$ is the projected area of the facial mask, $u_i$ is the local cartesian velocity vector, and $\Delta$ is the smoothed discrete delta function.

The shape of the mouth and its opening used in our simulations is shown in Fig. 1(a). The mouth geometry is 0.03 m wide and 0.01 m high. The face mask has asymmetrical curvatures around the face and its thickness at different locations around the face is heterogeneous with a mean thickness of 2 mm (Fig. 1(b,c)). According to past studies[16,18], a drag coefficient of 350 is used for the facial mask model to represent a typical non-medical grade facial mask. The human anatomy geometry and facial mask were generated using Blender (www.blender.org), an open-source software. Details of the numerical method and validation studies for jet-like flows, are documented elsewhere (Refs. 23-25).

## IV. Computational details

The computational domain includes a 4.0 m long, 1.0 m wide, and 2.5 m high room. The person is 1.83 m tall with the mouth located at an elevation of 1.67 m above the ground. The computational domain is meshed using a stretched Cartesian grid system with nodes clustered with a stretching ratio of 1.002 in all directions allowing for a resolution of 0.5 mm near the mouth. Hence, the computational domain involves over 0.6 billion grid nodes with grid spacing ranging in size from 0.5 mm in the vicinity of the face to nearly 5 mm at 4 m away from the face. A time step of $0.5 \times 10^{-3}$ s was selected to ensure a Courant-Friedrichs-Lewy (CFL) of $\leq 1.0$. We note that at this numerical resolution and for the range of Reynolds numbers during normal human breathing, the subgrid scale model in the LES is activated only in a small percentage of nodes in the vicinity of the mouth. Thus, our simulations resolve almost all relevant scales of motion and our work represents a high-fidelity simulation of human breathing.

Periodic boundary conditions are adopted in the spanwise direction. The no-slip boundary condition is used for the bottom boundary representing the ground, while at the top and outlet of the domain Neumann outflow boundary conditions are prescribed. The normal breathing cycle at the mouth is prescribed by imposing the streamwise velocity component, $u$ (m s$^{-1}$) using the following cosine function[32]:

$$u = \frac{V_t}{T_b A_{mo}} \pi \cos\left[\frac{2\pi t}{T_b} - \frac{\pi}{2}\right] \quad (6)$$

where $V_t$ is the tidal volume of breathing ($= 0.5 \times 10^{-3}$ m³), $T_b$ is the period of breathing (=5 s), $A_{mo}$ is the area of the mouth opening during breathing ($= 3.5 \times 10^{-4}$ m²), and $t$ is the time. The normal breathing waveform, which is prescribed at the mouth opening, is shown in Fig. 2.

Evaporation is known to impact the suspension and dispersion of saliva plume [6,17,33-36]. Saliva evaporation is a function of, ambient humidity, ambient temperature, and saliva particulate traveling velocity in the ambient air [17]; however, its rate has yet to be quantified. To consider the effect of evaporation on saliva particle transport, we employ an empirical approach by gradually reducing the exhaled saliva particle size and, consequently, gradually reducing their settling velocity. Our approach is inspired by the findings of Mittal et al. (2020) [17], who reported that once exhaled, saliva particles undergo a rather rapid evaporation process which leads to a reduction in their size. The rate at which we reduced the size of saliva particles in our simulations is shown in Fig. 3 and is based on the evaporation rate of saliva particles reported by Li et al. (2020) [33]. In this study, the exhaled air-saliva mixture (37 °C) consists of 10-μm saliva particles. As these 10-μm saliva particles enter the stagnant ambient air (20 °C), they undergo rather rapid evaporation and, after t ~ 0.2 s, their size reduces to 0.1 μm. At t > 0.2 s, their size is kept constant at 0.1 μm. Moreover, the saliva particle size range is selected to realistically represent saliva particulates produced during normal breathing, which varies between 0.1 μm and 10 μm (e.g., see Ref. 37). The settling velocity of saliva particles is calculated using the Stokes' law, as follows:

$$w_s = \frac{(\rho_0 - \rho_s) g d^2}{18\mu} \quad (7)$$

where $d$ is the size of the saliva particles and a function of time (see Fig. 3).

## V. Results and discussion

In this section we present and discuss the numerical simulations for two scenarios of normal breathing, with and without a facial mask. For both sets of simulations, the continuous spectrum of saliva particle sizes ranging from 0.1 μm to 10 μm is employed as described above. Both sets of simulations have been carried out on the same numerical mesh and using the same time step.

**VI. Normal breathing without a facial mask**

Fig. 4 shows instantaneous snapshots of the simulated saliva plume transport during normal breathing and without wearing a facial mask. For this case the simulation was continued for 90 s, a time during which the saliva plume was found to be still propagating forward, albeit at a progressively diminishing speed, due to the continuous and periodic exhalation process. Given the breathing period we consider herein, every 5 s a normal breathing cycle is completed effluxing the air-saliva mixture. As seen in Fig. 4, the ejected air-saliva mixture can be characterized as a longitudinal jet-like flow with a leading circular vortex ring. Without a face mask a single exhale generates enough momentum to propagate the air-saliva mixture up to 0.75 m away from the person (e.g., see Fig. 4(a) at t = 5 s and 10 s). The continuous exhalation process during normal breathing results in momentum accumulation near the front of the saliva plume. This accumulation of kinetic energy drives the forward motion of the saliva plume. As seen in Fig. 4(a), the saliva plume has reached a distance greater than 2.2 m away from the person in 18 breathing cycles, i.e., after t = 90 s of breathing. Given the temporal variation of saliva particle size due to evaporation (see Fig. 3), most of the saliva particles near the plume frontal boundary are 0.1 μm in size. The settling velocity of 0.1 μm saliva particles is about $3\times 10^{-7}$ ms$^{-1}$ (Eqn. 7) and, thus, under stagnant ambient air it would take days until they have fully settled down. Our simulation results show that within the first 57 s of continuous normal breathing, the saliva plume generated from normal breathing without wearing a facial mask can reach the CDC suggested social distancing length of 1.8 m.

The ejected air-saliva mixture into the stagnant indoor air induces an unstable shear layer whose rapid growth leads to the formation of a trailing jet and a corresponding leading vortex ring [38]. Given the normal breathing frequency in our study, every 5 s a new trailing jet with a leading vortex ring is generated. To better elucidate this process, we mark in Fig. 4 the leading vortex ring of the exhaled trailing jet generated at 'i-th' exhale cycle with $V_i$. The leading vortex ring propagates forward until it pinches off its trailing jet. This vortex pinch-off process[39] can be readily seen in Fig 4(a) (t = 10 s) where, vortex ring $V_1$ is seen to be entirely disconnected from its original trailing jet. Shusser and Gharib (2000) [40] further hypothesized that "*a vortex ring completes its formation and pinches off from its generating axisymmetric jet when the translational velocity of the ring becomes equal to the jet flow near the vortex ring.*" To better illustrate the vortex ring pinch-off mechanism in our simulation, we show in Fig. 5 the vorticity field of the leading vortex ring $V_1$ at two successive instants in times, which marks the initial stage of the pinch-off process [41-46]. As seen in this figure, at t=5 s, the vorticity field of the leading vortex ring is detached from that of the trailing jet setting the stage for the vortex ring pinch-off process.

The leading vortex rings, marked as $V_1$ to $V_{18}$ in Fig. 4(a), are circular and to better visualize their 3D structure we employ iso-surfaces of Q-criteria. In Fig. 6(a,b), we present snapshots of the simulated breathing flow vortical structures at t = 45.1 s and 61.2 s. We note that t = 45.1 s corresponds to a time 0.1 s after an exhalation event has started, and t = 61.2 s marks a time during an exhale when the air-saliva mixture velocity out of the mouth is near its peak. As seen in Fig. 6(a), the exhaled air-saliva mixture at t = 45.1 s generates an elliptical vortex ring very close to the mouth, which resembles the geometry of the mouth opening. Soon thereafter, however, and as the vortex ring propagates downstream, it begins to morph into an axisymmetric circular vortex ring – see for example the second vortex ring 0.62 m downstream which corresponds to the prior breathing cycle. Fig. 6(b) depicts the vortical structures of breathing at t = 61.2 s with a trailing jet connected to a leading vortex ring. In this figure, a pinched-off leading vortex ring, generated from the prior breathing cycle, can also be seen 0.7 m downstream of the mouth. Overall, the leading vortex rings seem to play a critical role in the spreading of saliva plume. In an experimental study using digital particle image velocimetry, Dabiri and Gharib (2004) [47] showed that the leading vortex rings contribute to mixing their content with the ambient fluid. More specifically, they reported that the leading vortex ring steadily grows in size and, by doing so, it entrains ambient fluid, which at some point makes up to 65% of the vortex ring volume [47]. We observed a similar mechanism in our simulation results (see, Figs. 4(a) and 6(a,b)), which show that the successive leading vortex rings play a crucial role in the mixing of the saliva plume of exhaled flow with the ambient air. Finally, our simulation results show that when the leading vortex rings reach about 0.75 m downstream, their vorticity field starts dissipating, while their remaining momentum accumulates and propels the vorticity field forward.

As the breathing continues, the newly formed leading vortex rings propagate and collide with vortex rings generated during prior breathing cycles (see Fig. 4). The exhaled air-saliva mixture during each breathing cycle accumulates kinetic energy near the saliva plume's frontal boundary. Over time, this process leads to slow but continuous propagation of the saliva plume forward. This incremental and periodic accumulation of momentum can be quantified using the temporal variation of the total kinetic energy of the normal breathing in Fig. 7. As seen, the total kinetic energy of the breathing without a mask undergoes two types of fluctuations: (1) a periodic increase/decrease due to the exhale/inhale cycles, respectively, and (2) a general increase due to the accumulation of energy, which is brought about by the successively generated breathing vortex rings. It can be readily seen that the total kinetic energy of normal breathing without a mask also undergoes a generally increasing trend, albeit at an asymptotically diminishing rate. The peaks of the periodic fluctuations of the total kinetic energy in Fig. 7 can be attributed to the air-saliva mixture's efflux during exhale cycles. The ambient fluid backflow into the mouth during the inhale cycles also produces a certain level of energy, which, however, is only a small fraction of the energy produced during exhales. The inhale-induced energy footprint can be seen as small kicks over the descending limbs of the total kinetic energy curve in Fig. 7.

**A. Normal breathing with a facial mask**

The instantaneous simulation results of the saliva plume transport with the facial mask are shown in Fig. 9. The facial mask has two main effects on the saliva plume transport during normal breathing. First, it re-directs the breathing jet flow slightly downward and, secondly, it significantly reduces the forward momentum of the breathing. Similar findings for other types of expiratory events are reported in Refs. 15 and 16. As a result of these effects, the facial mask seems to suppress the spreading of saliva plumes by dissipating its forward momentum. Our simulation results showed that, despite the continuation of normal breathing, after t = 102.5 s, the forward propagating momentum of the saliva plume at its frontal boundary reaches machine zero. The forward propagating of the saliva plume was halted at a time when the saliva plume has traveled a distance of 0.72 m, which is well below the CDC suggested social distance length of 1.8 m. The simulation was continued until t = 110 s, during which the saliva plume continued to somewhat spread in the vertical direction mainly due to the settling velocity of the saliva particles.

The effect of the facial mask on normal breathing vortex dynamics and saliva particle transport is illustrated in Fig. 6(c,d,e). As seen in Fig. 6(c), at t = 45.1 s, the only coherent flow structure in the flow field consists of an elliptical vortex ring, which is located in the space between the mouth and the facial mask. Since it has not yet been disrupted by the facial mask, this vortex ring is still intact and has enough energy to transport saliva particles downstream. We note, however, that in the case of breathing without the mask the leading vortex ring induced in the prior breathing cycle was still active 0.62 m downstream of the mouth (see Fig. 6(a)). In the breathing with the mask, on the other hand, the corresponding vortex ring is completely dissipated soon after passing through the facial mask. As vortex structures pass through the facial mask, deformation of the trailing jet and the leading vortex rings can be seen in Fig. 6 (d,e). As seen, the two vortical flow structures immediately downstream of the facial mask are asymmetrical due to the heterogenous thickness of the facial mask. As the vortical flow structures pass through the facial mask, they transport some saliva particles downstream until they start to dissipate soon after reaching about 0.15 m from the mouth. Hence, our results show that wearing a facial mask can prevent the generation of energetic periodic trailing jets and leading vortex rings and, consequently, suppress the forward propagation of the saliva particle plume.

The ability of a facial mask to dissipate the energetic vortical structures and suppress their forward momentum can be quantified by examining the time history of the total kinetic energy of the normal breathing in Fig. 7. As seen, the total kinetic energy of the normal breathing flow with the facial mask is roughly one order of magnitude smaller than that of the same breathing flow without a facial mask. Khosronejad et al. (2020) [16] reported a similar finding for other expiratory events. As seen in Fig. 7, the total kinetic energy of the breathing with the facial mask undergoes periodic fluctuations with two local peaks. The higher peaks correspond to the forward motion of air-saliva mixture during exhale cycles. In comparison, the smaller peaks show the kinetic energy that is generated by the backflow of ambient fluid during inhale cycles (Fig. 8(c,d)). Unlike normal breathing without the mask, the total kinetic energy of the normal breathing with the facial mask does not exhibit the overall increasing trend and only undergoes periodic fluctuations about a mean value. Therefore, our results provide strong evidence that a facial mask can effectively dissipate the translational energy of the breathing flow and, therefore, prevent the plume of saliva particulates from traveling farther than 0.72 m away from the mouth. They also show that a facial mask essentially halts the forward propagation of the saliva plume at a distance well below the CDC social distancing guidelines.

## VII. Conclusion

We conducted high-fidelity numerical simulations of normal breathing to investigate the vortex dynamics and transport mechanisms of saliva particle plume under indoor conditions with and without a facial mask. The size range of saliva particles in our study was in the range of 0.1 μm to 10 μm, and the saliva evaporation's effect was incorporated by decreasing the saliva particle size over time. Our findings regarding the rate and distance of saliva propagation with and without the facial mask are succinctly summarized in Fig. 10, which depicts the time variation of the traveled distance of the saliva plume frontal boundary along the streamwise direction. We find that the initial speed of the saliva plume propagation is relatively high for both cases with and without the facial mask. While soon after the first breathing cycle is completed (i.e., t = 5 s), the slope of both curves starts to decline. Unlike the case of normal breathing with the facial mask, however, for which saliva propagation is rapidly halted after 102.5 s when the front of the plume reaches 0.72 m downstream away from the mouth, breathing without a mask gives rise to a plume that continues to propagate downstream, albeit at a gradually decreasing speed, even after 90 s of simulation time. In fact, for this latter case, saliva particles exceed in about 57 s the CDC suggested social distancing length of 1.8 m and continue to propagate downstream. Our findings suggest that the CDC guidelines for social distancing can be effective only if they are used in combination with facial mask wearing.

**Acknowledgments**

This work was supported by grants from the National Science Foundation (EAR-0120914) and a sub-award from the National Institute of Health (2R44ES025070-02). Also, the authors would like to thank Christian Santoni and Kevin Flora for their contributions to the development of the mask and human anatomy models, respectively.

**Data Availability Statement**

The data that support the findings of this study are available from the corresponding author upon reasonable request.

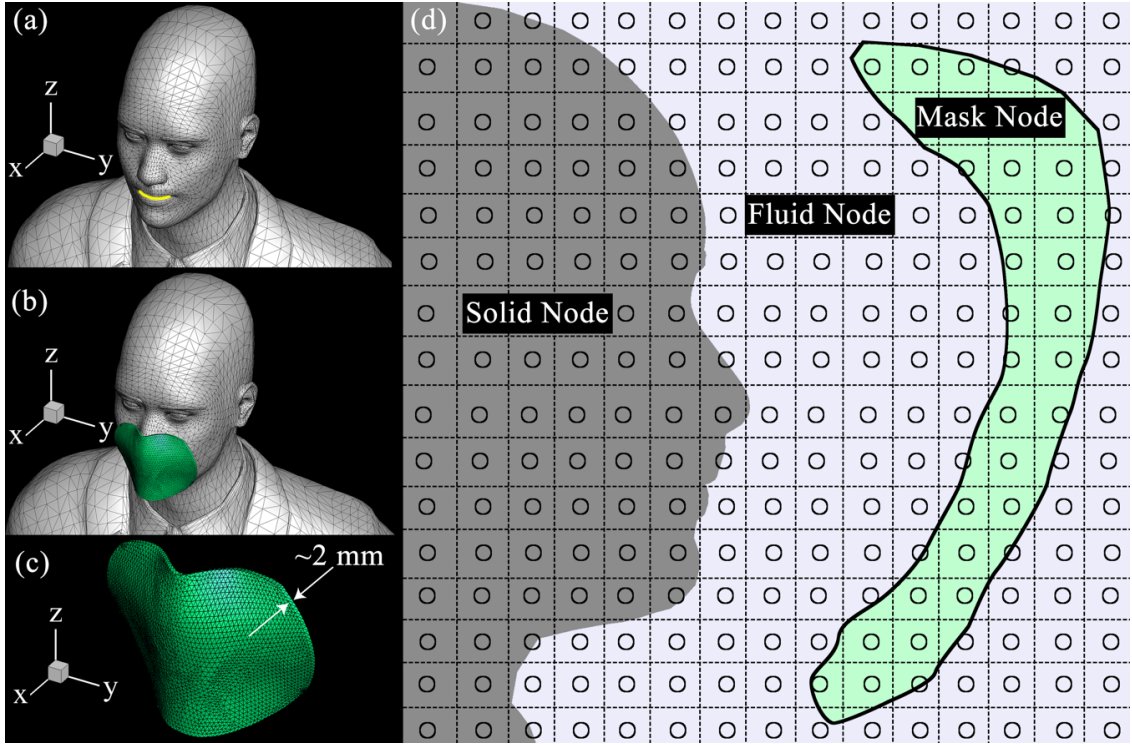

**Figure 1:** Simulated human anatomy and the facial mask. (a) illustrates the face and the mouth opening in yellow. (b) shows the human anatomy and the facial mask covering the face and mouth. (c) shows the facial mask. (d) is a schematic of the background grid system from side view depicting the solid nodes inside human anatomy (dark gray), fluid nodes in the air (light gray), and mask nodes inside the mask (green). Black triangles in (a) to (c) illustrate the unstructured grid system we employed, in the context of CURVIB, to discretize the human anatomy and facial mask.

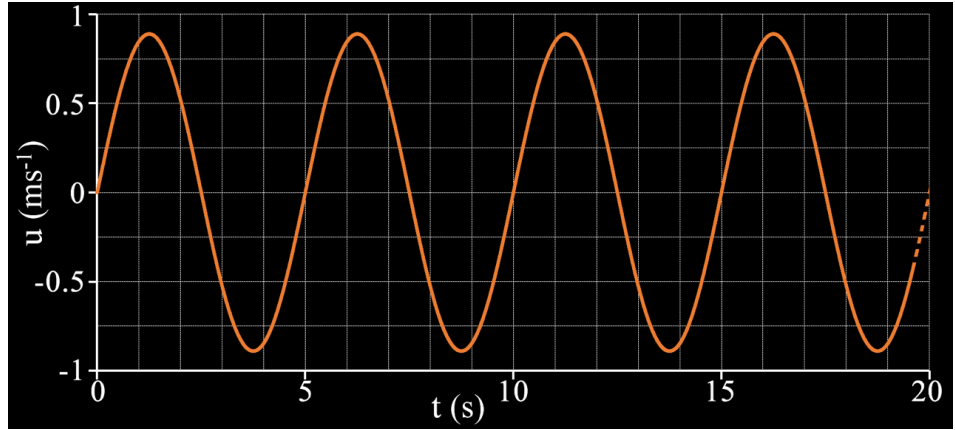

**Figure 2:** The instantaneous normal breathing waveform for the streamwise velocity component, u, imposed at the opening of the moth. The waveform is computed using the cosine function in Eqn. (6). The positive and negative values of u represent exhale and inhale cycles, respectively.

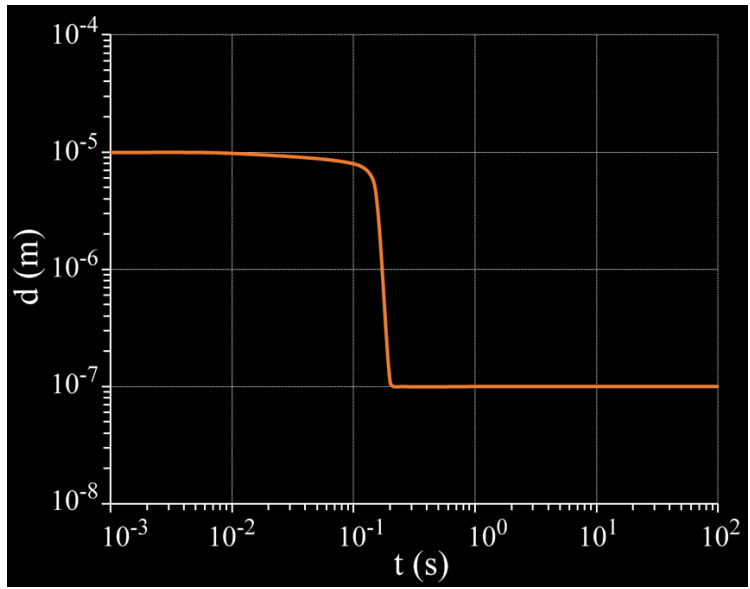

**Figure 3:** Time variation of saliva particle size, d, as prescribed in the model to incorporate the effect of saliva particles' evaporation.

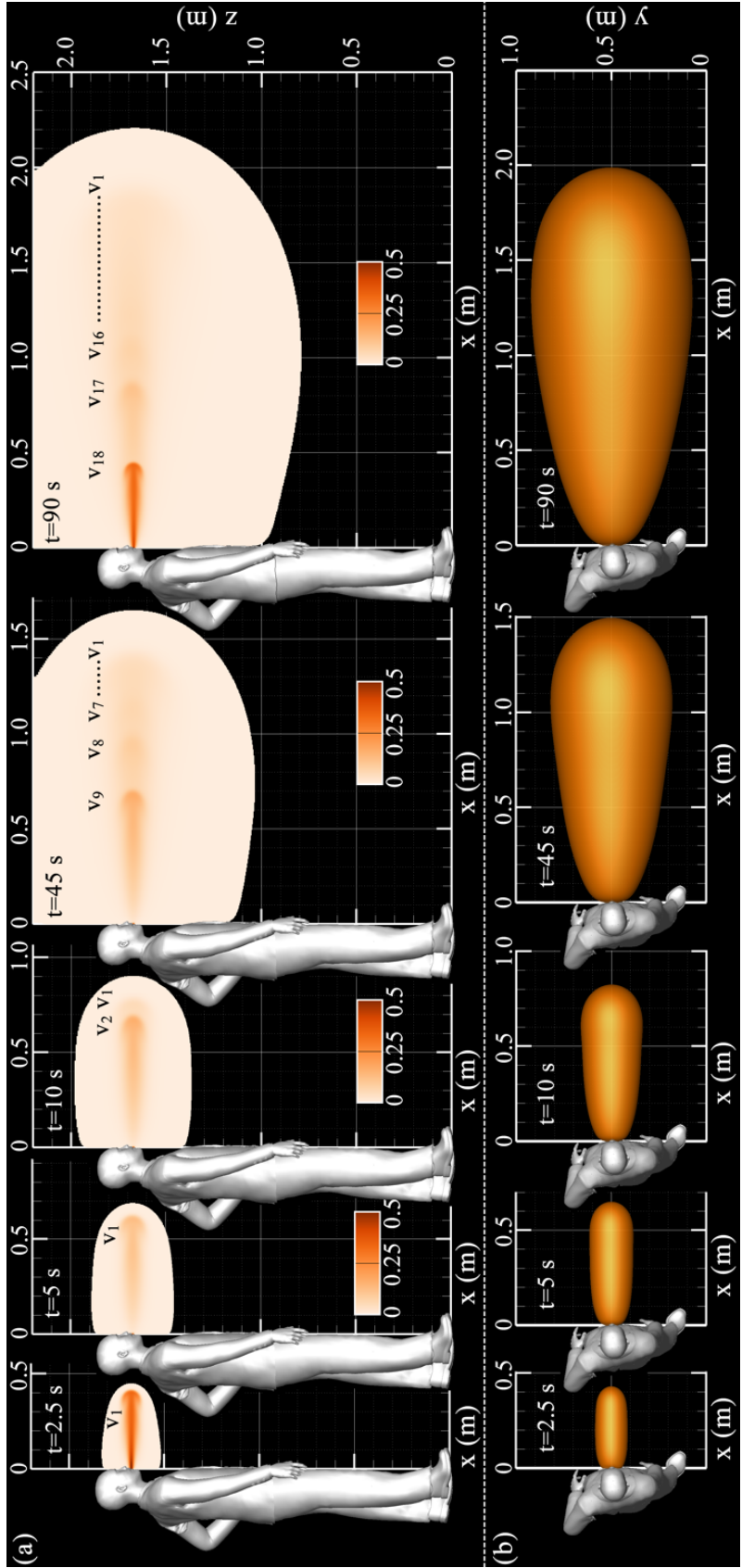

**Figure 4:** Simulated snapshots of saliva plume concentration (volume fraction) (a) contours and (b) iso-surfaces during normal breathing without a facial mask. Inserts of (a) and (b) are shown on the sagittal plane and from top view, respectively. As seen in (a), the exhaled air-saliva mixture generates periodic leading vortex rings which are marked as $V_i$ in (a), where 'i' shows the 'i-th' leading vortex ring created at 'i-th' breathing cycle.

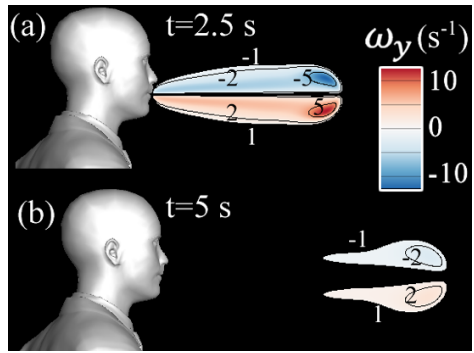

**Figure 5:** Simulated instantaneous out-of-plane vorticity (s$^{-1}$) contours during normal breathing without a facial mask on the sagittal plane. (a) marks the time when the first exhale is completed, while (b) marks the end time of the first breathing cycle when the inhale is completed. The vorticity field in (a) shows the trailing jet, and the leading vortex induced by the exhale during the first breathing cycle. The beginning of the pinch-off process, i.e., vortex ring disconnection from its trailing jet, can be seen in (b).

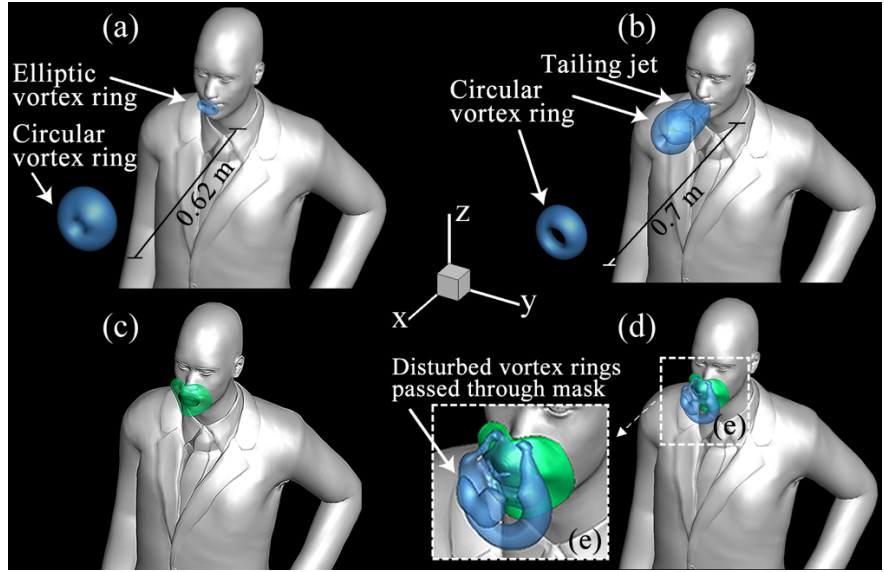

**Figure 6.** Simulated evolution of the normal breathing vortical flow structures without (a,b) and with (c,d,e) the facial mask in 3D. (a,c) and (b,d) show the simulated vortex rings after t = 45.1 s and 61.2 s, respectively. The vortexes are visualized using the iso-surfaces of Q-criterion (=0.1). (a) shows the formation of an elliptic vortex ring immediately outside the mouth and a circular vortex ring at 0.62 m downstream. (b) depicts the vortical structures of breathing during the exhale, marking the formation of a trailing jet connected to a leading vortex ring, plus a pinched-off vortex ring from the prior breathing cycle at 0.7 m ahead of the mouth. The only coherent vortical structure present in (c) is an elliptic vortex ring formed in the space between the mouth and the facial mask – rendered visible by making the mask translucent. (d) and (e) show the breathing flow vortical structures deformed and re-directed by the facial mask, whereas their counterparts' undisturbed vortical structures are shown in (b).

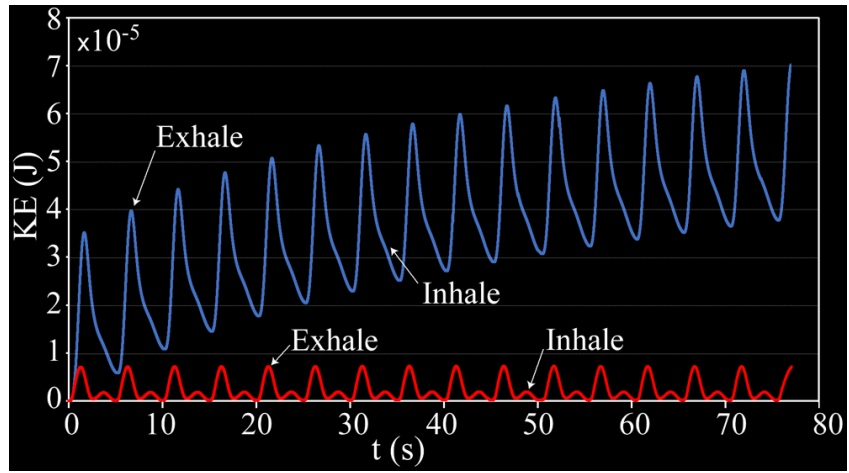

**Figure 7:** Time history of the total kinetic energy (J) of the normal breathing with (red line) and without (blue line) the facial mask. The total kinetic energy of the breathing with the facial mask is one order of magnitude smaller than that without the facial mask.

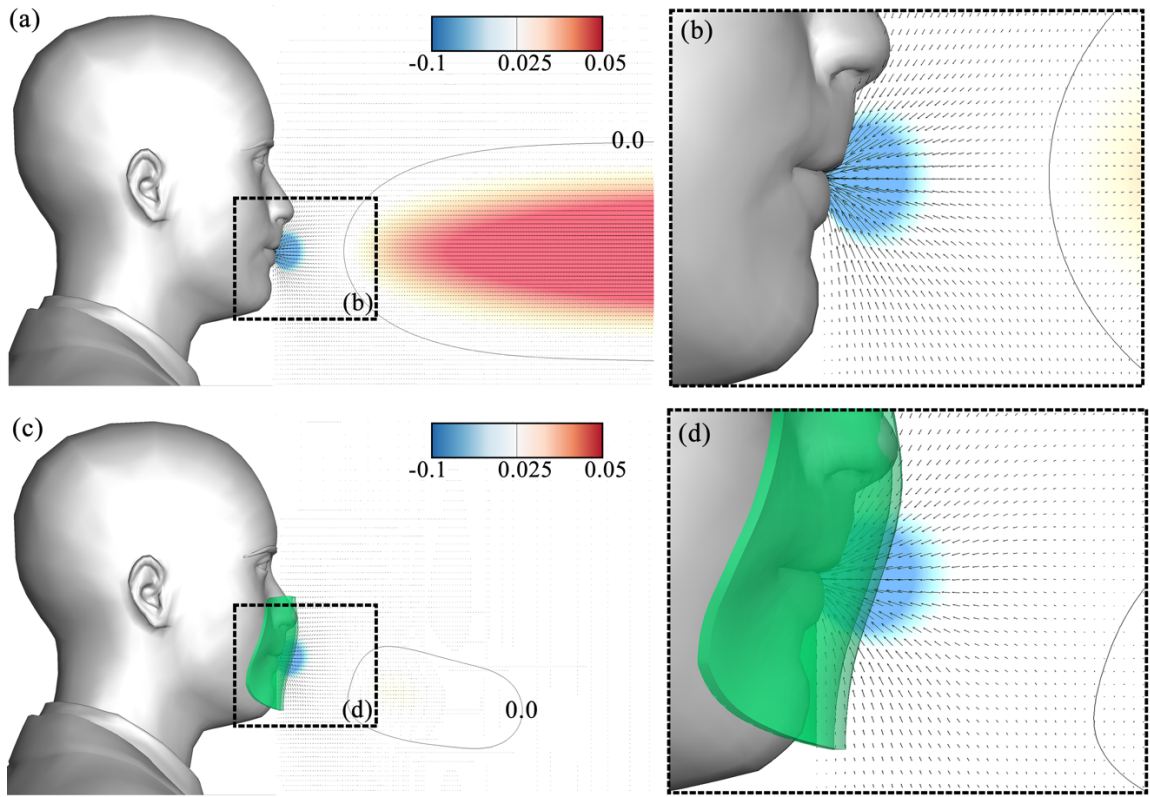

**Figure 8:** Simulated instantaneous velocity vectors superimposed over the contours of streamwise velocity component (in ms⁻¹) during the normal breathing without (a,b) and with (c,d) the facial mask on the sagittal plane. The blue regions near the mouth mark the air-saliva mixture's backflow into the mouth during the inhale cycle at t = 3.7 s. For the sake of clarity, only one out of eight velocity vectors are .

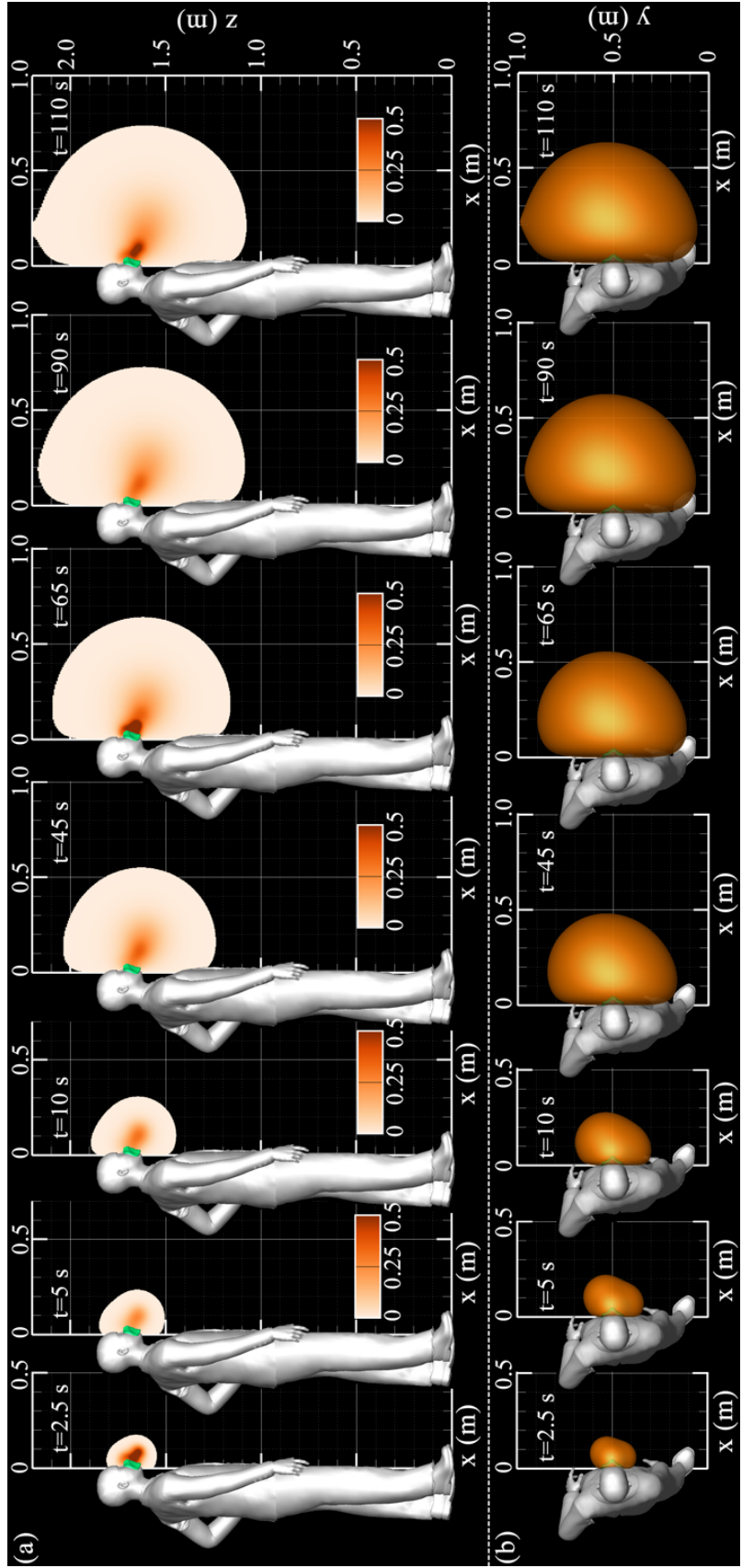

**Figure 9:** Simulated snapshots of saliva plume concentration (volume fraction) (a) contours, and (b) iso-surfaces during normal breathing with a non-medical-grade facial mask. Inserts of (a) and (b) are shown on the sagittal plane and from the top view, respectively. The iso-surfaces of saliva plume in (b) are slightly asymmetrical due to the heterogeneity of the face mask's thickness.

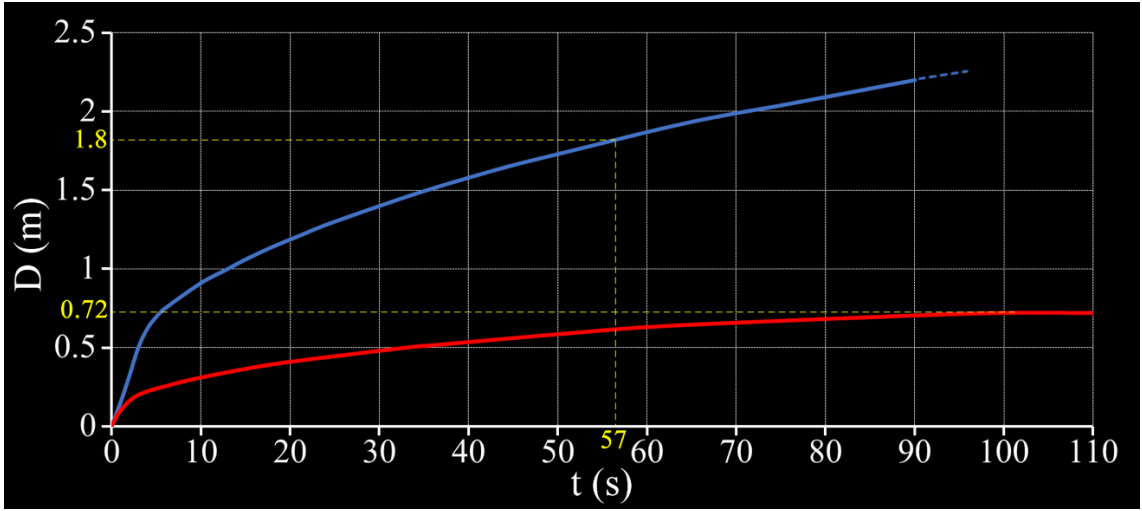

**Figure 10:** Simulated temporal variation of the spreading length (D) of saliva particle plume in streamwise direction during the normal breathing with (red line) and without (blue line) the facial mask. Blue line shows that in normal breathing without the face mask, it takes 57 s until the saliva plume reaches 1.8 m away from the person in normal breathing without the face mask. Unhindered with the facial mask, after 100 s of breathing, the saliva plume continues to propagate froward and beyond 2.2 m away from the person. The red line plateaus after 102.5 s, when the saliva particle plume has reached 0.72 m away from the person, demonstrating the facial mask's effectiveness to limit the spreading of the saliva plume during normal breathing.